\documentclass[12pt]{article} 
\usepackage{natbib}
\usepackage{graphicx,color}
\usepackage{subcaption}
\usepackage{amsmath}
\usepackage{amssymb}
\usepackage{adjustbox}
\begin{document}
\title{Supergranular Fractal Dimension and Solar Rotation\\}
\author{Sowmya G  M $^1$ \thanks{sowmyabharth5@gmail.com}, Rajani G $^2$, U Paniveni $^3$, $^4$, R Srikanth $^4$}
\maketitle
$^1$ GSSS Institute of Engineering and Technology for Women, Mysuru-570016, Karnataka, 
India\\
$^2$ PES College of Engineering, Mandya - 571401, Karnataka, India.\\
$^3,^4$ Bangalore University, Jnanabharathi, Bengaluru – 560056\\
$^4$ Poornaprajna Institute of Scientific Research, Devanahalli, Bangalore-562110, Karnataka, India\\

\maketitle
%

\begin{abstract}
We present findings from an analysis of the fractal dimension of solar supergranulation as a function of latitude, supergranular cell size and solar rotation, employing spectroheliographic data in the Ca II K line of solar cycle no. 23. We find that the fractal dimension tends to decrease from about 1.37 at the equator to about 1 at 20 degree latitude in either hemisphere, suggesting that solar rotation rate has the effect of augmenting the irregularity of supergranular boundaries. Considering that supergranular cell size is directly correlated with fractal dimension, we conclude that the mechanism behind our observation is that solar rotation influences the cell outflow strength, and thereby cell size, with the latitude dependence of the supergranular fractal dimension being a consequence thereof.
\end{abstract}



\section{Introduction}
 Supergranules, the large convective eddies discovered by Hart in the year 1950 and later characterized by \citet{leighton1960rb} are  believed to be visible manifestations of sub-photospheric convection currents. Typically, these cellular patterns have a horizontal flow velocity in the range of 0.3 to 0.4 km/s, an autocorrelation length scale of around 30 Mm and a lifetime of about 24 hour \citep{simon1964velocity}. The supergranular pattern as a whole tends to be irregularly surface filling  \citep{leighton1962velocity} and has an estimated lifetime of about 2 days \citep{gizon2003wave}. While their horizontal flows may reach ~300-400m/s, their upflows are an order of magnitude slower. Unlike granules, they are not thought to be truly convective, which explains why they are better observed in Dopplergrams than in intensitygrams.  Indeed, this is a reason why they were initially discovered through Doppler images. It is known that supergranular cell boundaries coincide with the chromospheric networks, attributed to magnetic fields flushed to the cell boundaries by the horizontal flow (Simon and Leighton, 1964). The size and flow spectrum associated with supergranulation include smaller cells in such a way that the spectrum of supergranules leads to the spectrum of granulation \citep{hathaway2000photospheric} and has a dependence on Solar cycle phase and total irradiance \citep{mandal2017association}.
 Here it may be noted that both Doppler signals and the spectral component due to granules are visible in SDO/HMI data \citep{williams2014analysis}.
 
A number of researchers have noted the effects of interaction between solar activity and the supergranular magnetic network. Based on an analysis of spectroheliograms spanning seven consecutive solar maxima, \citet{singh1981dependence} claim that the chromospheric network cell size is smaller at the solar maximum phase than at the solar minimum phase. This is in consonance with the findings of \citet{kariyappa1994variability} on the chromospheric network variability, of \citet{berrilli1999average} on the network geometry, and of \citet{raju2002dependence}, who study magnetic field influence on network scale, but differs from a study on the related velocity and magnetic fields \citet{wang1988relationship}, and \citet{munzer1989pole}, who have reported larger network cells areas in higher magnetic activity regions.   
 
 The supergranular rotation rate at the solar equator has been reported by various authors and found to be about 3$\%$ more than the surface plasma's rotation rate, a phenomenon termed as 'supergranular superrotation' \citep{duvall1980equatorial,  beck2000supergranulation}, but it should be noted that this is probably a projection effect and not a genuine wave phenomenon \citep{hathaway2006supergranule}. Based on a time-distance helioseismology analysis of the SOHO-MDI, the pattern of supergranulation is found to be oscillatory \citet{gizon2003wave}, generating waves with a time period between six and nine days. The apparent superrotation may be explained by the fact that the waves are largely prograde.     	
 
 The fractal dimension is a useful mathematical representation for describing the complexity of geometrical structures and for understanding the underlying dynamics \citep{mandelbrot1975geometry}.An object is called a fractal if it displays self-similarity at different scales. Fractal analysis has been used to study the turbulence of the magnetoconvection of solar magnetic fields \citep{lawrence1993multifractal,  stenflo2003flux}. Fractal analysis has been used in the context of solar surface studies, such as in the context of dopplergrams \citep{meunier1999fractal} and Ca II K filtergrams of SoHO MDI \citep{paniveni2011solar} and of KSO data \citep{chatterjee2017variation, rajani2022solar}. The fractal nature of supergranulation was studied in detail by \citet{paniveni2005fractal} and its relation to solar activity by \citet{paniveni2010activity}, where the role of turbulence on the complexity of the cell was indicated.  Pic du Midi data was to calculate the granulation pattern's fractal dimension \citep{roudier1986structure}, which was the first application of fractal dimension investigation to a solar surface phenomenon. For smaller granules, they obtained a fractal dimenson of  2 for large granules and $1.25$ for smaller ones. 
   \citet{berrilli1998geometrical} used fractal analysis to explain the turbulent origin of supergranulation. They chose an intensity threshold and produced binary image representing the chromospheric network and used a medial axis transform (skeleton) of the binary image to unleash the geometrical properties of the cells. To calculate the degree of circularity of supergranular cells, \citet{srikanth2000distribution} used the tessellation method on the supergranulation pattern. 
 
 \section{ Data and Analysis}
\begin{figure}
\centering
\includegraphics[width=70mm]{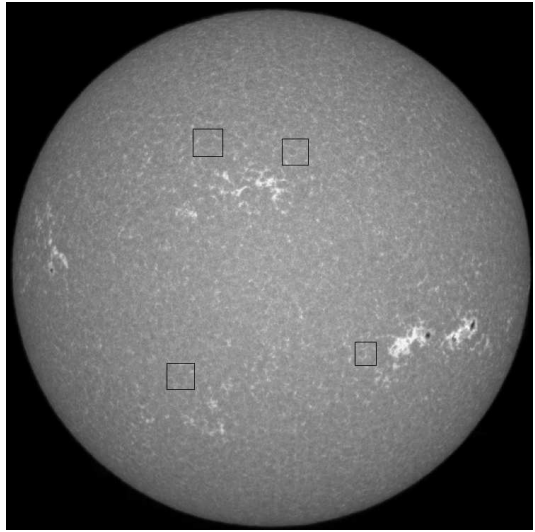}
\caption{Spectroheliogram of Ca II K from KSO, indicating supergranules selections, taken during cycle no. 23, in particular the active phase of October, 2000. The image orientation is N-S.}
\label{fig:Fulldisk}
 \end{figure}

 This analysis uses the quiet region data (in both quiescent and active phases) of the solar cycle no. 23 (covering the years 1996  to 2008) from the Kodaikanal Solar Observatory (KSO)\footnote{https://kso.iiap.res.in} archives. Figure \ref{fig:Fulldisk} depicts data obtained during the active phase of this cycle. The KSO's dual telescope is equipped with a Ca II K spectroheliograph with a spectral disperson of 7 $\AA$/mm near 3930 $\AA$. It employs a 6 cm image obtained with a  Cooke photovisual triplet of 30 cm, onto which sunlight is reflected by a 460 mm diameter Foucault siderostat. Light with a band with of 0.5 $\AA$ is admitted by the exit slits. The images are suitably time-average to remove the effects of p-mode oscillations. 

Well-formed supergranular cells within an angular distance of $20^\circ$ are selected by visual inspection, where the restrictionis made to minimize projection effects, cf. \citet{rajani2022solar}. Figure \ref{fig:Fulldisk} is part of a full-disk image in which we highlight a few regions where we are able to visually identify well-defined cells.
 Per day the setup generates 144 images  with post-averaged time cadence of 10 min. As the image resolution is 2 arcsec, which is twice the granular scale, it is expected that our results are insensitve to granular effects.  About 400 well-defined cells were extracted from quiet regions within the belt between $20^\circ$ N and S. The area-perimeter relation is obtained from them forms the basis for deriving the fractal dimension \citep{paniveni2018latitudinal}. 

\begin{figure}
	\centering
	\begin{subfigure}{0.75\textwidth}
	\includegraphics[width=80mm]{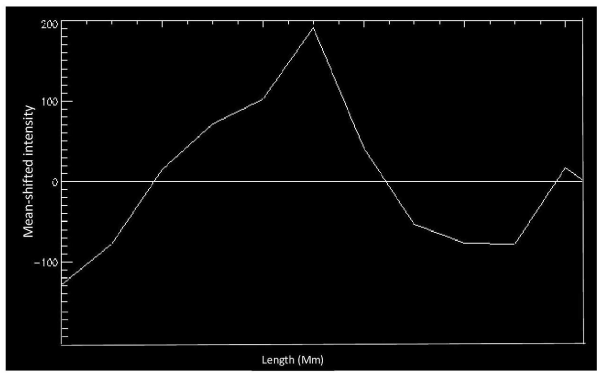}
	\end{subfigure}
	\caption{Ca II spectroheliogram scan: mean-shifted profile of a selected supergranule showing two crests, which stand for the cell boundary. If the peak position was ambigous, one could potentially try to use a Gaussian profile to fit the cell wall. However, as in the above case, the position could be unambiguously determined. The cell area and perimeter are obtain with multiple such scans. (The negative values corresponds to points with below-mean intensity.)}
	\label{fig:intensity scan}	
\end{figure}
 
The methodology was manual and not automated. It goes briefly as follows: first the visually identified cell subjected using IDL software to a ``two-dimensional tomography'', i.e., multiple sequential scans, such as shown in Figure ~\ref{fig:intensity scan}. In each scan, the cell boundaries define the area included in the scan, which is added to obtain a consolidated area, while the locus of boundaries across scans determines the cell perimeter. 

Our analysis, based on direct visual inspection, yields a cell size in consonance with other works which employ methods that track individual cells \citep{paniveni2005fractal,hagenaar1997distribution}. The latter reference infers cell diameter between 13 to 18 Mm, employing a tessellation procedure based on the steepest gradient algorithm, obtained a characteristic cell diameter in the range 13-18 Mm, which is half of the cell scale obtained using methods such as autocorrelation method or spherical harmonics decomposition \citep{hathaway2000photospheric}. The cause of this discrepancy is a matter under current investigation, to be reported elsewhere.
 
\begin{figure}
	\centering
\includegraphics[width=80mm]{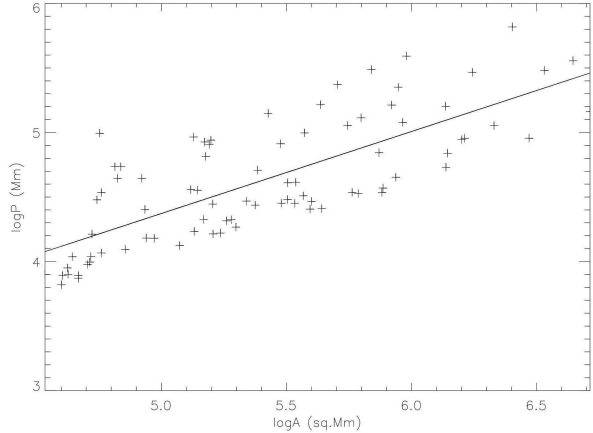} 
\caption{Log-log plot of supergranular perimeter vs area in units of Mm and Mm$^2$, respectively. The displayed data consists of about 130 cells, corresponding to the first data point in Figure \ref{fig:FD}.}
\label{fig:A-P}
\end{figure}

Figure~\ref{fig:A-P} gives the area vs perimeter plot for the analyzed cells, demonstrating a power-law relationship. If $P$ and $A$ denote the cell's perimeter and area, respectively, then the fractal dimension $D$ is obtained according to:
\begin{equation}
\begin{aligned}
\centering
D\delta \log(A) = 2\delta \log(P).
\label{eq:fractal}
\end{aligned}
\end{equation}
Perfect circles or squares, for which the area increases quadratically as a function of the perimeter, we find that the fractal dimension $D = 1$. The more the cell structure deviates from regularity by being denticulate (i.e., the boundaries are craggy and rugged), the more it causes greater perimeter length to enclose a given area, and thereby the more is the increase of the fractal dimension towards 2. 

The chosen region of study, which is about 30$^\circ$ subtended about the image center, should contain approximately 300 cells per image. Thus, in principle, a greater number of cells can be employed than used in this study. In automated methods of cell extraction (such as the steepest-gradient method based tessellation technique of one of the authors here (e.g., \citet{srikanth2000distribution}), or autocorrelation based extraction of cell scale (e.g., \citet{raju1998dependence}, by the same authors), a greater region can be mechanically covered for study. However, such methods require a degree of interpretation, such as (in the former case) whether the extracted cells are precisely supergranules or include other cell-like regions of smaller or larger scale. In the latter case, the autocorrelation scale may be enhanced as an artefact of open cells, which lack a well-defined boundary. The present manual method has the advantage of visually selecting well-defined cells, but being time-consuming, yields fewer cells in a given time. Further, the present method may involve a selection effect in that it may be biased towards cells of smaller size. This is because apparently they tend to be better defined than larger cells, which tend to have more broken / diffuse boundary walls. Ideally, it would be apt to develop a supervised machine-learning algorithm that is trained by the present visual inspection method.

\section{Result and Discussion}
 \subsection{ Cell size, fractal dimension and rotation} 
\begin{figure}
	\includegraphics[width=3in]{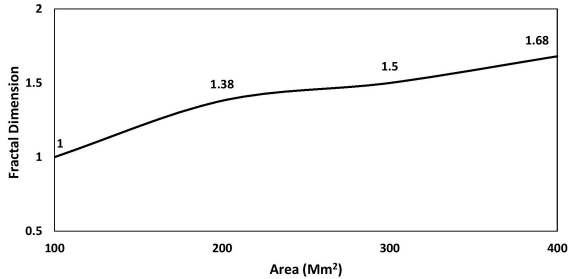}
 \caption{Supergranular fractal dimension dependence on area, showing that larger cells are more irregular shaped. Here, the area parameter is grouped into bands of size 100 Mm$^2$, which is large enough to enable inclusion of a statistically significant number of data points. In passing, it may be mentioned that this behavior is in agreement with the findings of Meunier (1999) for active regions.}
\label{fig:FD}	
\end{figure}

   In a first analysis, we look at how the fractal dimension varies with supergranular length scale. We have considered four size ranges, combining data across all latitudes. Figure  ~\ref{fig:FD} depicts a broad but well-established dependence of fractal dimension on the area of the supergranular cells and is shown in different ranges of the supergranular cell area. For those cells whose area is below 100 Mm$^2$, the fractal dimension is found to be about 1, meaning that they are quite regular in shape. On the other hand, for an area between (100-200) Mm$^2$, (200-300)  Mm$^2$, (300-400)  Mm$^2$ the fractal dimension is found to be about 1.38, 1.5 and 1.68 respectively indicating a more irregular-shaped perimeter.
  
The choice of a band of 100 Mm$^2$ to classify the cells is rather arbitrary, but found to be convenient for our data set. Thus our main observation here is that the smaller supergranular cells are more regular shaped than larger ones. 
This agrees with the result reported by \citet{srikanth2000distribution}, who find that larger cells have less regular boundaries (quantified through a ``circularity'' parameter). This feature is attributed to the idea that supergranular outflows become choppier at larger distances, reflected in the irregularity of the swept-out magnetic fields. Here, it is of interest to note that \citet{berrilli1998geometrical} used fractal analysis to explain the turbulent origin of supergranulation.

 \subsection{Latitude, Solar rotation and fractal dimension}
  In a second analysis, the fractal dimension is computed for the latitude belts (0-3), (3-6), (6-9), (9-12), (12-15), (15-18) and (18-21), the data comprise cells from both hemispheres. In this case, cells are not sifted according to size. Columns $\#$2 and $\#$3 of Table~\ref{tab:latitude} gives the latitude range and corresponding fractal dimension. It shows that at lower latitudes, the estimated fractal dimension is higher than that at the higher latitudes (cf. \citep{raju2020asymmetry}). The result is given in Table ~\ref{tab:latitude}. The data of Figure \ref{fig:FD} and Table \ref{tab:latitude} together suggests that supergranular cell sizes fall slightly at higher latitudes in the selected belt, in agreement with the observation of \citet{raju1998dependence}.
    
  \begin{table}[htb]
 \begin{tabular}{|c|c|c|c|}
\hline
&  Latitude range & Fractal Dimension & Rotation \\
Sl.No. & $\theta$ in degree & & Rate\\
& & & $\Omega$/2$\pi$ \\
\hline
1 & {0-3} & $1.37 \pm 0.03$ & 461.6\\
 		\hline		
 	2& {3-6} &1.3 $\pm 0.02$ &461\\
 		\hline
 	3&{6-9}&	$1.23 \pm 0.02$ &	460\\
 		\hline
 	4&	{9-12}&	$1.2 \pm 0.01$ &	459\\
 		\hline
 		5&	{12-15}&	$1.16 \pm 0.02$ &	457.8\\
 		\hline
 		6&{15-18}&   $1.1 \pm 0.01$ &455.9\\
 		\hline
 		7&	{18-21}&	$1 \pm 0.01$&	453.5\\
 		\hline
 	\end{tabular}
 \caption{The individual perimeter-vs-area plots are used to obtain fractal dimension for each latitude belt, plotted in Figure~\ref{fig:A-P}. For each latitude belt, the fractal dimension derived is based on about 50-100 cells.}
 \label{tab:latitude}
 \end{table}
   The latitudinal dependence of supragranular fractal dimension suggests a connection to solar differential rotation and possibly to supergranular superrotation. The cellular rotation rate, as determined  by \citet{hathaway2012supergranules}, is:
\begin{equation}
\begin{aligned}
\centering
\Omega (\theta,\lambda)/2\pi=   [1+g(\lambda)](454- 51\sin^2\theta  - 92 \sin^4\theta),
\label{eq:rot-theta}
\end{aligned}
\end{equation}
where $\lambda$ is latitude and $g(\lambda) = \tanh(\lambda/31)[2.3-\tanh((\lambda-65)/20)] /73.3$ is expressed in Mm and g($\lambda$) is a dimensionless quantity. With a typical value $\lambda = 32$ Mm, the value of $(1+g (\lambda))$ turns out to be about 1.017. The value of rotation for the mid-belt is given in Table~\ref{tab:latitude}.
Eq. (\ref{eq:rot-theta}) shows that the rotation rate falls off as one moves away from the equator in either hemisphere, similar latitudinal dependence of the fractal dimension.

\begin{figure}
	\centering
\includegraphics[width=90mm]{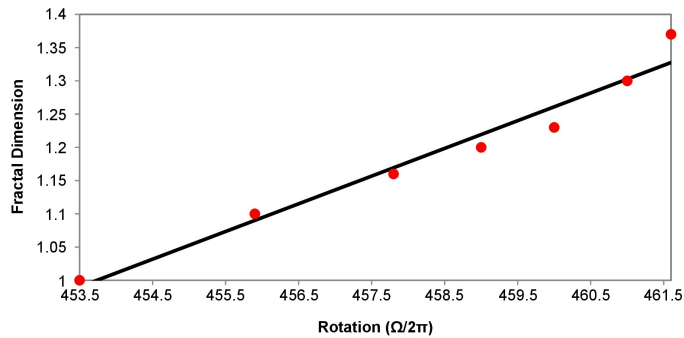} 
\caption{Variation of fractal dimension with rotation from the data of Table \ref{tab:latitude} and Eq. (\ref{eq:rot-theta}). The linear fit comes from assuming a linear relation between the two variables, and requiring a best fit subject to the constraints of Eqs. (\ref{eq:rot-theta}) and (\ref{eq:key}).}
\label{fig:rotation}
\end{figure}

In order to connect the observation given by Eq. (\ref{eq:rot-theta}) to our data, we shall assume a simple linear relation between fractal dimension and rotation given by $D = a + b(\Omega/2\pi)$, for certain real parameters $a$ and $b$. The form of Eq. (\ref{eq:rot-theta}) leads us to the relation
\begin{equation}
\begin{aligned}
\centering
D = 1.34 - 3.5 \sin ^2 \theta - 6.3 \sin^4\theta.
\label{eq:key}
\end{aligned}
\end{equation}
which is found to provide a reasonable fit to the data of Table \ref{tab:latitude}. 
Using Eqs. (\ref{eq:key}) and (\ref{eq:rot-theta}) to eliminate $\theta$, we obtain: 
\begin{equation}
\begin{aligned}
\centering
D = -29.6 + 0.067(\Omega/2\pi)
\label{eq:Dw}
\end{aligned}
\end{equation}
plotted in Figure~\ref{fig:rotation}. 
Other slightly different versions of the dependence Eq. (\ref{eq:rot-theta}) are possible, e.g., \citep{korzennik1989seismic}, and accordingly we may obtain slight variations of Eq. (\ref{eq:Dw}).

Two causes may be at play working hand in hand to produce the rotational  dependence of fractal dimension, given by Eq. (\ref{eq:Dw}). First is that, as we reported above, cell sizes fall towards higher latitudes (Table \ref{tab:latitude}), which may be a rotational effect and can be understood as follows. The differential rotation through the dynamo action causes an enhancement of quiet sun magnetic fields at higher latitudes. This field enhancement is expected to have a constricting influence on cell size \citep{singh1981dependence}, leading to smaller cells at higher latitudes, as confirmed by \citet{raju1998dependence}. And as we show later (below in Eq. (\ref{eq:deltal})), larger cells are expected to have a greater fractal dimension. By virtue of Eq. (\ref{eq:rot-theta}), we know that rotation speed falls towards higher latitudes.  These considerations provide a basis for the observed direct correlation between $D$ and the rotation rate. Another possible cause is related to the fact that when the radial outflow of a supergranule encounters the ambient plasma at the cell boundary,  the fluidic stress and hence turbulence is expected to be relatively less where the plasma rotation speed is lower, assuming uniform outflow speed across the latitudes.  Correspondingly, the cell boundaries at latitudes associated with slower rotation, namely the higher latitudes, are expected to be less corrugated, or in other words, have lower fractal dimension, as we find in Table \ref{tab:latitude}.

\section{Conclusion \& Discussions}
We have found that the fractal dimension for supergranulation is directly correlated with supergranular cell size (Figure~\ref{fig:FD}), but anti-correlated with latitude (Table ~\ref{tab:latitude}). Taking into account the observed quartic polynomial relationship between Solar rotation and the sine of the latitude, Eq. (\ref{eq:rot-theta}), we have proposed a simple dependence of fractal dimension on solar rotation. We now briefly and qualitatively consider the question of a potential underlying mechanism to explain this behavior and that we hope to understand more quantitatively in a future work.

The latitude dependence of fractal dimension $D$ is expected to be influenced by its dependence on the scale of supergranulation and the quiet Sun magnetic field distirbution.  We now discuss the nature of these two dependences. With regard to the latter, we remark that the magnetic flux tubes, ``frozen'' into the plasma, have the constricting property, essentially because charged particles aren't allowed to cut across field lines.This is due to the Lorentz force,  given by $\vec{F}_{\rm L} \propto \vec{v} \times {\bf B}$, where ${\bf B}$ and $\vec{v}$ represent magnetic field intensity and velocity, respectively.  Indeed, the flow of plasma across a field line is forbidden in the limit of extremely high electrical conductivity because it would generate enormous eddy currents \citep{alfven1942existence}. 

We now speculate on a potential qualitative scenario that can account for our results. Assuming that supergranules are convective cells, magnetic field is expected to be accumulated at the supergranular edges thanks to the above magnetohydrodynamical feature. Larger number of flux tubes transported to the edges of the larger cells due to convective motions and the associated solar rotation  may be a key factor in determining how strongly the supergranular outflow pushes against the ambient plasma, resulting in smaller cells at higher latitudes in the chosen latitudinal range.  

Since the cell wall is formed by a heating of the overlying plasma by the magnetic flux swept by the supergranular convective flow, larger cells typically show more fluctuations and discontinuities in the cell wall, and hence larger fractal dimension. This may explain the direct correlation between cell size and the fractal dimension (Figure \ref{fig:FD}). 
We propose a simple model that tries to capture the above idea. For the turbulent medium described by Kolmogorov theory applied to Solar convection associated with supergranulation, we expect the relation between the horizontal speed $v_{\rm horiz}$ and the cell size $L$ being given by:
\begin{equation}
v_{\rm horiz} = \eta^{1/3} \times L^{1/3},
\label{eq:vhoriz}
\end{equation}
where $\eta$ is connected to the plasma injection rate \citep{paniveni2004relationship}.  Letting $T = L/v_{\rm horiz}$ represent the time that a plasma fluid element takes to traverse from the point of upflow at the cell center to the boundary, and $\delta_{\rm horiz}$ represent the standard deviation in the horizontal velocity, we may then estimate that the standard deviation induced in $L$ is given by
\begin{equation}
\delta_L = T \delta_{\rm horiz} = \eta^{-1/3}L^{2/3}\delta_{\rm horiz},
\label{eq:deltal}
\end{equation}
which implies that the cell boundary has greater spread, the greater is the cell size. \citet{paniveni2004relationship} estimate using SOHO dopplergram data that $\eta, \delta_{\rm horiz}$ and the mean value of $L$ are, respectively, $2.89 \times 10^{-6}$ km$^2$ s$^{-3}$, 74.1 m/s and 33.7 Mm. Substituting these values into the right hand side of Eq. (\ref{eq:deltal}), we obtain about 5.4 Mm for $\delta_L$, which is close to the value of standard deviation in $L$ of 8.96 Mm reported by \citet{paniveni2004relationship}. 

It is not unreasonable to assume that the standard deviations mentioned above obtained over many cells also indicate the variation of the corresponding variables over different times and positions in a given cell. Under this assumption, Eq. (\ref{eq:deltal}) can be interpreted as asserting that the boundaries of larger cell show greater fluctuation, and thus by extension, greater fractal dimension, consistent with the plot in Figure \ref{fig:FD}. Our result appears to support previous studes \citep{srikanth1999chromospheric,srikanth2000distribution}, which reports that larger cells have a more craggy perimeter.

\citet{raju1998dependence} have reported a decrease in the autocorrelation scale of supergranules as one moves to higher latitudes until $\pm 20^\circ$, and an increase thereafter until $\pm 30^{\circ}$. In conjunction with Figure \ref{fig:FD}, this would suggest that the fractal dimension must have an analogous latitude dependence, with minima around $\pm 20^\circ$. Thus, whilst $D$ has the expected behavior at the lower latitudes, it appears that other factors must be invoked to explain its behavior farther up. Here we note that quiet Sun fields are reported to show enhancements around the equator and $\pm 30^{\circ}$ \citep{harvey1998Solar}. This, in light of the preceding argument, would be consistent with the data of Table \ref{tab:latitude}, except that we would expect a dip in $D$ close to the equator. In conclusion, it appears that the latitude dependence of $D$ that we find is the resultant of the somewhat conflicting constraints imposed by the cell scale and quiet Sun magnetic field distribution. We may conclude that further study, using a different method of cell statistics analysis to process a larger number of cells, is needed to unravel the detailed behavior of $D$ as a function of latitude.

It will of be of interest to try to quantitatively obtain Eq. (\ref{eq:Dw}) based on these consideration, which would then lead to Eq. (\ref{eq:key}) in conjunction with Eq. (\ref{eq:rot-theta}). In future works, we propose to return to the same data, but using other approaches, such as an autocorrelation, spectral analysis or an automated tessellation. 

Here it is worth noting that a turbulent origin of supergranulation has been studied, and in particular \citet{berrilli1998geometrical} have used fractal analysis in this context. In the theory of turbulent energy cascade, the  Kolmogorov spectrum for energy as function of wave number $k$ is given by $k^{-\frac{5}{3}}$  implies that the variance of temperature varies with length scale as $r^{2/3}$, while variance of pressure varies as $r^{4/3}$ \citep{paniveni2005fractal}. \citet{mandelbrot1975geometry} showed that the fractal dimension of an isosurface is given by $D = D_E - 2 \times \langle \zeta \rangle$, where $D_E$ is the Euclidean dimension of the object (here 2, for supergranulation) and $\langle \zeta\rangle$ is the exponent in the functional form of variance for the given quantity. Accordingly, for isotherms and isobars we find $D=5/3 \approx 1.66$ and $D = 4/3 \approx 1.33$, respectively. Our data in Table \ref{tab:latitude} show that each latitude, the fractal structure of supergranulation is closer to an isobaric than isothermal pattern. It would be interesting study whether the assumed linear behavior that underlies Eq. (\ref{eq:Dw}) is related to this.

\section{Acknowledgement}
We thank Indian Institute of Astrophysics (IIA) for providing Ca-K filtergram data, and Fiaz for providing technical help with image handling. We are grateful to Prof. J. Singh for his valuable suggestions and support.

\bibliography{./References}
 \end{document}